# The Open Vault Challenge
## *Learning how to build calibration-free interactive systems by cracking the code of a vault*

**Jonathan Grizou**
Center for Research and Insterdicisplinarity (CRI)
Université de Paris, France
jonathan.grizou@cri-paris.org

## Abstract

This demo takes the form of a challenge to the IJCAI community. A physical vault, secured by a 4-digit code, will be placed in the demo area. The author will publicly open the vault by entering the code on a touch-based interface, and as many times as requested. The challenge to the IJCAI participants will be to crack the code, open the vault, and collect its content.

The interface is based on previous work on calibration-free interactive systems that enables a user to start instructing a machine without the machine knowing how to interpret the user's actions beforehand. The intent and the behavior of the human are simultaneously learned by the machine.

An online demo and videos are available for readers to participate in the challenge. An additional interface using vocal commands will be revealed on the demo day, demonstrating the scalability of our approach to continuous input signals.

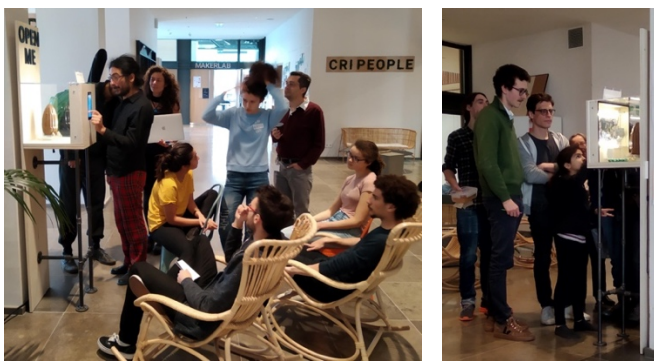

*Figure 1: Left: staffs and students trying to crack the code. Right: 9-year old girl showing off her new skill.*

## 1 Introduction

When giving instructions to a machine, humans are funneled into how to provide those instructions by the interface we present to them. Buttons are to be pressed and the letters printed on them indicate what they do. Such interfaces act as translators from the raw signals of the user to the symbolic level that the machine work on. In the case of a remote control, we transform the human action of pressing the button marked with a 1 into a digital signal that instructs the machine to display the channel 1 on the screen.

In other scenarios, we may want to use more complex signals than discrete button pressing events from a user. This might be vocal commands, gestures, or even brain signals for severely disabled persons. In such cases, the signal to meaning mapping needs to be learned because no explicit rules can be hand-crafted by the engineer. This is the well-known process of training a classifier that can be used to predict the class/meaning of a user's signal.

Training such classifiers requires to have access to a database of labeled samples. In the brain-computer interaction case, we need samples of brain signals (e.g. EEG) from a human and their associated meaning (e.g. a binary feedback yes/no). However, acquiring such dataset requires us to have access to the ground truth label, that is the intention of the user.

This raises a chicken-and-egg problem. To infer the intent of one user while using an interface, we first need to be able to decode its communicative signals. But to train such a decoder, we initially need to know what the user intent is in order to collect a labeled training dataset. In practice this is solved by explicitly instructing the user what to do during a, often tedious, calibration session.

A curious mind might ask: Is there a way out of this chicken-and-egg problem? Can we design calibration-free interfaces that adapt to the specificity of each user on the fly?

The vault challenge, described in the next section, has been designed to guide the participants into solving this chicken-and-egg problem by themselves.

## 2   Vault Challenge

Opening a vault is always done via an interface, for example via a key or a code. Our vault is just the same, we enter a 4-digit code using a touch-screen. But instead of pressing directly on numbers, the user reacts to the information displayed on the screen by the vault.

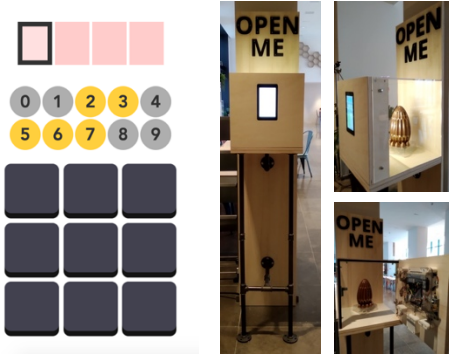

*Figure 2: Left: User interface on the vault. Right: Photo of the physical vault containing a massive chocolate egg as a reward.*

Gifts, in the form of food items, will be placed in the transparent vault. The participants will be able to watch and record the author as he enters the code. From there, participants will have to reverse engineer the algorithm powering the interface, crack the code, open the vault, and collect the gift.

As a demo for readers, we present two levels below. It should be emphasized that there is no prior information shared between the user and the vault, except, of course, that the user knows the correct 4-digit code. All that is shown in the videos linked below is enough to infer the correct code.

**A taster for beginners**
In this video https://youtu.be/TcmF19BWrVk, a user enters the correct code. Can you find out what the code is? You can check if you are correct by entering the code again at http://openvault.jgrizou.com/#/ui/level_1.json.

This level is relatively simple. The vault highlights some numbers in yellow and others in gray. As a user, if the number you are looking at is yellow, you press the yellow button, respectively gray otherwise. By elimination, the vault can infer the digit the user has in mind.

**The calibration-free challenge**
In this video https://youtu.be/iipCFJAcU88, a user enters a different code. Can you find out what the code is? You can check if you are correct by entering the code again at http://openvault.jgrizou.com/#/ui/level_4.json.

This level is much more intriguing already, the buttons have no colors. How are we supposed to know if a button is associated with yellow or gray? This is the calibration-free challenge solved by the underlying algorithm.

To visualize the simultaneous identification of the code and the colors of the buttons, you can enter any code at http://openvault.jgrizou.com/#/ui/level_3_visible_nocheck.json. For example, enter 0000, while considering in your mind that the central button is yellow, and all the others are gray. Then try again by reversing the colors.

Additional levels show the use of continuous signals. e.g. at http://openvault.jgrizou.com/#/ui/level_5_visible_nocheck.json, users can click on arbitrary 2D locations on the screen, instead of pressing buttons.

**User engagement.** This challenge has been tested in our institute, students and staff were continually engaged over many days, see Figure 1. We expect this demo to produce a similar effect at IJCAI given the form and nature of the challenge. Advanced levels are challenging enough for experienced researchers and will make them discover an intriguing approach to interactive systems.

**Hardware.** The vault is a cube made of wood and acrylic measuring 40x40x40cm. If for any reason the physical vault does not reach the destination, the demonstration will become a screen-and-mouse-based. In both cases, the demo will be simultaneously available online.

## 3   Discussions

The calibration-free challenge has previously been explored in the field of human-robot interaction [Lopes et al., 2011; Grizou *et al.*, 2013] and brain-computer interfaces [Kindermans et al., 2012; Grizou *et al.*, 2014a]. See [Cederborg, 2017] for a detailed review.

The solution implemented for this demo consists of analyzing the consistency of a user wrt. all possible objectives he might have [Grizou, 2014]. The context allows to generate a set of hypotheses about the user's intent. Each hypothesis allows assigning a set of labels to the data received by the user. For N hypothesis, we create N different datasets with the same set of signals but different labels. Finding the hypothesis with the most consistent labeling can inform us both about the intent of the user (the most likely hypothesis) and the signal-to-meaning mapping used by the user (a classifier that can predict the meaning of a user's signal).

This problem does not map directly to the usual supervised, unsupervised, or reinforcement learning paradigms. We believe it could be beneficial to researchers of various fields to be exposed to this challenge and get a taste of what can be achieved, which might lead to applications in their own research field. For example, the planning under uncertainty problem is of particular interest [Grizou *et al.*, 2014b].

Finally, our live demo will demonstrate that, with our calibration-free approach, the vault can be opened with speech commands using arbitrary words in any language and without any prior calibration.

## Acknowledgments

We thank the CRI for the fellowship allowing developing this demo. Many thanks to Edwin Paquiot and Joanna May Lee for their countless design advice, as well as to students and staff at CRI for testing this setup extensively.

## References

[Lopes et al., 2011] Manuel Lopes, Thomas Cederborg, and Pierre-Yves Oudeyer. *Simultaneous acquisition of task and feedback models*. International Conference on Development and Learning (ICDL), 2011.

[Grizou *et al.*, 2013] Jonathan Grizou, Manuel Lopes, andPierre-Yves Oudeyer. *Robot learning simultaneously a task and how to interpret human instructions.* International Conference on Development and Learning and Epigenetic Robotics (ICDL), 2013.

[Kindermans et al., 2012] P.-J. Kindermans, D. Verstraeten, and B. Schrauwen. *A bayesian model for exploiting application constraints to enable unsupervised training of a P300-based BCI.* PloS one, 2012.

[Grizou *et al.*, 2014a] Jonathan Grizou, Inaki Iturrate, Luis Montesano, Pierre-Yves Oudeyer, and Manuel Lopes. *Calibration-Free BCI Based Control.* International Conference on Artificial Intelligence (AAAI), 2014.

[Cederborg, 2017] Thomas Cederborg. *Artificial learners adopting normative conventions from human teachers*. Paladyn, Journal of Behavioral Robotics, 2017.

[Grizou, 2014] Jonathan Grizou. *Learning from Unlabeled Interaction Frames.* PhD Thesis, 2014.

[Grizou *et al.*, 2014b] Jonathan Grizou, Inaki Iturrate, Luis Montesano, Pierre-Yves Oudeyer, and Manuel Lopes. *Interactive Learning from Unlabeled Instructions.* International Conference on Uncertainty in Artificial Intelligence (UAI), 2014.